\documentclass[12pt]{article}

\usepackage{amsmath}

\usepackage{amssymb}

\usepackage{array}

\usepackage{graphicx}

\usepackage{color} 

\usepackage{multicol}

\usepackage{lscape}

\usepackage{geometry}
\usepackage[labelfont=bf,labelsep=period,justification=raggedright]{caption}

\usepackage[sort&compress, numbers, square]{natbib}
\renewcommand{\cite}[1]{\citep{#1}}

\makeatletter
\renewcommand{\@biblabel}[1]{\quad#1.}
\makeatother

\date{}

\newcommand{\Mol}{{\mathcal M}} 

\newcommand{\aSetA}{A}   
\newcommand{\aSetC}{C}   

\newcommand{\Rea}{{\mathcal R}} 
\newcommand{\aRea}{\rho}        

\newcommand{\LHSset}[1]{\mathrm{LHS}(#1)}        
\newcommand{\RHSset}[1]{\mathrm{RHS}(#1)}        

\newcommand{\reactsto}{ \to}   

\newcommand{\GCL}{G_{CL}} 

\newcommand{\RED}[1]{{#1}}

\newcommand{\TF}{\mathit{TF}}
\newcounter{defcount}          
\newenvironment{definition}[1]{\addtocounter{defcount}{1}\vspace{3mm}\textbf{Definition \thedefcount: [#1]}\it}{\vspace{2mm}}

\newcommand{\url}[1]{}
\newcommand{\doi}[1]{}

\begin{document}

\begin{flushleft}
{\LARGE
\textbf{Molecular Codes in Biological and Non-Biological Reaction Networks}
}
\\ \vspace{0.3cm}
Dennis G\"{o}rlich$^{1,2,\ast}$, 
Peter Dittrich$^{1,2,\dagger}$, 
\\ \vspace{0.3cm}
$^1$ Bio Systems Analysis Group, Jena Centre for Bioinformatics (JCB) and Institute of Computer Science,
Friedrich-Schiller-University Jena, D-07743 Jena, Germany
\\
$^2$ Jena School for Microbial Communication (JSMC)
\\
$^\ast$ E-mail: dennis.goerlich@uni-jena.de
$^\dagger$ E-mail: peter.dittrich@uni-jena.de \\ \vspace{0.3cm}
Version: 25. May 2011
\end{flushleft}

\section*{Abstract}
Can we objectively distinguish chemical systems that are able to
process meaningful information from those that are not suitable for
information processing? Here, we present a formal method to assess the
semantic capacity of a chemical reaction network.
The semantic capacity of a network can be measured by analyzing the
capability of the network to implement molecular codes.
We analyzed models of real chemical systems (Martian atmosphere
chemistry and various combustion chemistries), bio-chemical systems
(gene expression, gene translation, and phosphorylation signaling
cascades), as well as an artificial chemistry and random networks.
Our study suggests that different chemical systems posses different semantic
capacities. 
Basically no semantic capacity was found in the atmosphere chemistry of Mars and
all studied combustion chemistries, as well as in highly connected random
networks, i.e., with these chemistries molecular codes cannot be
implemented. High semantic capacity was found in the bio-chemical
systems, as well as in random networks where the number of second
order reactions is at the number of species. 
\RED{Hypotheses concern
  the origin and evolution of life}. We conclude that our
approach can be applied to evaluate the information processing
capabilities of a chemical system and may thus be a useful tool to
understand the origin and evolution of meaningful information, e.g.,
at the origin of life.

\section*{Introduction}
In recent years great advances have been made in understanding the
bio-chemical basis of biological information processing. 
For theoretical analysis of biological information Shannon's theory of
communication \cite{Shannon1948} has  been applied very successfully  
in various domains, like genomics \cite{Schneider1990}, bacterial
quorum sensing \cite{Mehta2009}, or signaling in molecular systems
\cite{Lenaerts2008}. 
The mathematical theory of communication focusses on uncertainty of events and
intentionally neglects semantic aspects of information, because
``{\it they are irrelevant for the engineering
  problem}'' (Shannon \cite{Shannon1948}, p. 1).
However, in order to obtain a full understanding of biological
information, studying also semantic as well as pragmatic aspects would
be important, if not necessary \cite{Monod1971,Kueppers1990}. 
Although syntax, semantics, and pragmatics are interdependent,
as detailed in the $Co^3$ approach \cite{Tsuda2009}, we concentrate here on semantic aspects 
of molecular networks in order to keep our formalism and analysis
clear and concise.
 
In general, semantics refers to the relation between a sign and its 
meaning and can be described by a code. An example is the genetic
code, which is a mapping between codons (signs) and amino acids
(meanings) \cite{Barbieri2008}. An important
property of this mapping or relation is its contingency, that is, 
the relation could be different and thus is not determined by the signs and
meanings alone \cite{Monod1971,Barbieri2008}. We say that the relation
between signs and meanings is contingent, if the relation cannot be
derived by applying natural laws to the signs and meanings alone. This
implies that by natural laws we can only derive the relation by
knowing in addition a context under which the signs are interpreted.
Furthermore it implies that there is potentially another context under
which the signs are interpreted differently.

It is thus sometimes stressed that the relation between signs and meanings
cannot be explained by physical laws
\cite{Pattee2001}, like the natural laws do not
help in understanding the written law or the grammar of a language. However, more often than not
this notion of independence from natural laws is the cause for
confusion. 
So, in order to properly use
semiotic concepts in biology, we should provide a proper -- ideally formal -- link
from these concepts to the realm of physics. To achieve this
we take the following strategy:
(Step~1) Select an experimentally grounded and reliable formal description of the targeted
  biological system. Here, we take the reaction
  network as this formal description.
(Step~2) Provide precise, not necessarily formal, definition of the semiotic concepts
  that shall be applied to the system. Here, we take the notion of an
  organic code as reviewed by Barbieri~\cite{Barbieri2008}.
(Step~3) Interpret these definitions by linking
 them to the formal description of the biological system. Here, this
 is done by our formal definition of a molecular code.
With this a semiotic concept gets -- at least
partially -- operationalized by means of physical experiments.
In particular, it allows us to incorporate contingency
in a formal model of molecular codes.

To illustrate the basic idea of an explicit modeling of contingency we
will briefly discuss an example reaction network, which exhibits a contingency. Figure \ref{fig:examplenet}A
shows a reaction network containing eight molecular species
$\{A,B,C,D,E,F,G,H\}$ and four reactions. Here, we assume that the
network contains all possible reactions that can appear when mixing
these molecules. The network then is a complete model of the world,
i.e., no species and reactions are missing that are physically
possible.
A \textit{mapping} in a reaction network relates molecular species. Here, for
example, $\{A\}$ can be mapped on $\{C\}$ by reaction $A+E \rightarrow
E+C$. $\{E\}$ is necessary for the reaction to happen and thus we call
it a \textit{molecular context}.  The network can implement a \textit{molecular code}, if
there exists a set of molecular species that can be mapped on a
second set of molecular species in at least two different ways.
In this example network the sets  $S=\{A,B\}$ and $M=\{C,D\}$ fulfill this
property. $S$ (\textit{domain}) maps on $M$
(\textit{codomain}) by applying the context $\{E,H\}$. No two elements of
the domain $S$ map to the same element in the codomain $M$. There exist an
alternative molecular context $\{F,G\}$ which realizes a different mapping
between domain and codomain. The existence of these two alternative
mapping qualifies these as codes, since they emerge from a
contingency.

\section*{Materials and Methods}

In this section we provide a formal definition of a molecular code as a contingent mapping
that can be realized by a reaction network, then we formally define the semantic
capacity of a reaction network based on the number of molecular codes
it can realize, and finally describe two algorithms for finding all
molecular codes of a reaction network.

\subsection*{Definition of Molecular Code}
\label{sec:molec-organ-codes}

\begin{figure*}[!ht]
\centering
\includegraphics[width=0.6\textwidth]{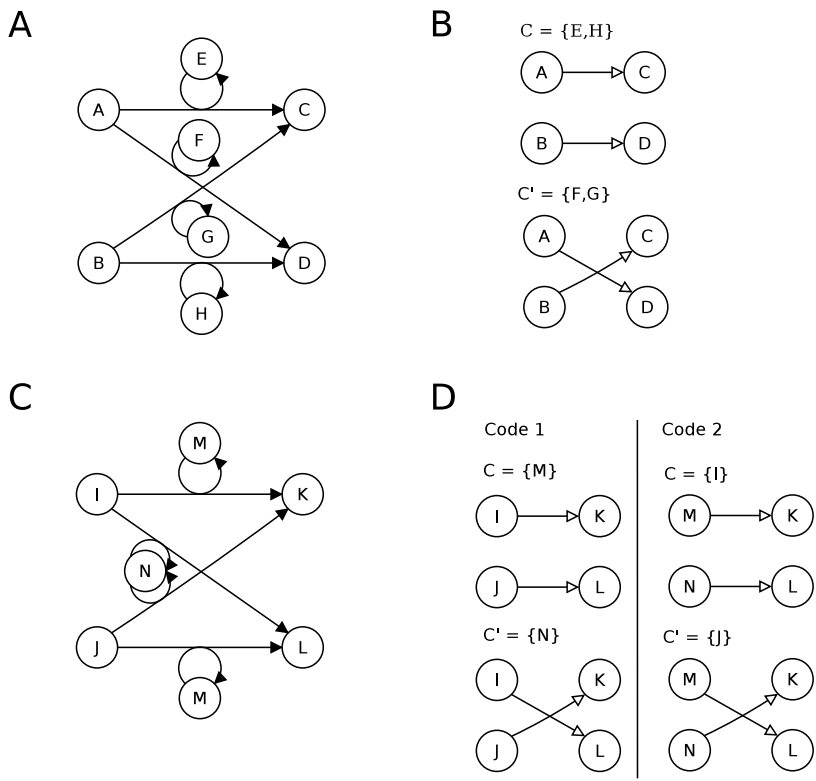}
\caption{{\bf Two exemplary reaction networks containing molecular codes.}
Panel A: Chemical reaction network $\langle \Mol, \Rea
  \rangle$ with species $\Mol = \{A, B, C, D, E, F, G, H \}$ and
 reaction rules $\Rea = \{ A + E \reactsto C + E, \dots \}$;
panel B: Code pair that can be realized by the network in panel A. The
binary molecular codes are characterized by $S = \{A, B \}$, $M = \{C, D
\}$, and the two codemakers $C=\{E,H\}$, and $C'=\{F,G\}$;
panel C: Chemical reaction network with species $\Mol = \{I, J, K, L,
M, N, M \}$. The two species labeled 'M' denote the same species;
panel D: Two molecular code pairs can be realize by the network in panel
C. 
}
\label{fig:examplenet}
\end{figure*}

A \emph{reaction network} $\langle\Mol, \Rea \rangle$ is defined by
a set of molecular species $\Mol$ and a set of reactions $\Rea$ 
occurring among the molecular species $\Mol$. 
See Figure~\ref{fig:examplenet}A for an example.
For each reaction $\aRea \in \Rea$, let $\LHSset \aRea$
and $\RHSset \aRea$ denote the set of reactant species
(left hand side) and product species (right hand side) of
reaction $\aRea$, respectively. 

A subset of molecular species $\aSetC \subseteq \Mol$ is 
called \emph{closed}, iff the application of all possible reactions
from $\Rea$ on $\aSetC$ does only produce species from $\aSetC$, i.e.,
for all $\aRea \in \Rea$ with $\LHSset \aRea \subseteq \aSetC$:
$\RHSset \aRea \subseteq \aSetC$ \cite{fontana_buss_94}. 
For any set of species $\aSetA \subseteq \Mol$ there exists a smallest
closed set $G_{Cl}(A)$ containing $\aSetA$ \cite{ped:SDZB2000}. 
We say that $G_{Cl}(A)$ is the \textit{closure} of $\aSetA$. 
Intuitively, the closure of a set of species contains all those
species that can be reached by arbitrary long reaction pathways
among the species of that set.

\begin{definition}{molecular mapping}
Given a reaction network $N = \langle \Mol, \Rea \rangle$ and
two sets of molecular species  $S, M \subseteq \Mol$,
we say that $f: S \to M$ is
a \emph{molecular mapping} with respect to the reaction network $N$, iff there exist
a set of species $C \subseteq \Mol$ (called context), such that
for each $s, s' \in S$ with $s \neq s'$: 
 $f(s) \in \GCL(C \cup \{ s \})$ and 
 $f(s') \notin \GCL(C \cup \{ s \})$. 
If there is a molecular mapping $f$ with respect to $N$, we
also say that $N$ can \emph{realize} the molecular mapping $f$.
\end{definition}

Note that in a reaction network
there is usually more than one molecular context $C$
that realizes a particular molecular mapping $f$.
Intuitively, in order to ``compute'' $f(s)$ with the reaction network
$N$, we put all molecules from the context $C$ together with $s$
in a reaction vessel. Then we repeatedly apply  all applicable 
reaction rules and add
the products to the reaction vessel until no novel molecular species
can be added anymore. Then we check which molecular species 
from $M$ is present, which must be -- according to our definition --
unique and the result of $f(s)$.

\begin{definition}{molecular code}
  Given a reaction network $N=\langle \Mol, \Rea \rangle$ and a
  non-constant\footnote{A mapping $f: S \to M$ is called non-constant,
    iff there exists $s, s' \in S$ such that $f(s) \neq f(s')$.}
  molecular mapping $f: S \rightarrow M$, with $S,M \subseteq \Mol$
  with respect to $N$, we call the mapping $f$ a 
  \emph{molecular code} with respect to $N$, if all other
  mappings $g: S \to M$ with the same domain $S$ and codomain $M$  like $f$ can also be realized by the reaction network
  $N$.
\label{def:mc}
\end{definition}

The definition catches the notion of contingency as described above,
i.e., the elements of the domain can be mapped to the elements of the codomain
in an arbitrary way by changing the context.
In order to keep our study tractable, we will focus on 
molecular codes that are binary, i.e., where $S$ as well as
$M$ contain exactly two molecular species \cite{Goerlich2009}. 
We will also not study molecular
mappings that are only partially contingent, here.
For binary molecular codes our definition can be
reformulated more explicitly:

\begin{definition}{binary molecular code (BMC)}
Given a reaction network $\langle \Mol, \Rea \rangle$ and
two binary sets of molecular species 
$S=\{s_1,s_2\} \subseteq \Mol$ and 
$M = \{m_1, m_2\} \subseteq \Mol$. 
The mapping $f: S \rightarrow M$ 
is called  a \emph{binary molecular code}, iff there exist two sets $C \subseteq
\Mol$ (called codemaker) and $C' \subseteq \Mol$ (called alternative
codemaker) such that the following conditions hold:
\begin{eqnarray*}
f(s_1) \in \GCL(\{s_1\} \cup C) \textnormal{, and } f(s_2) \notin
\GCL(\{s_1\}   \cup C ) \textnormal{, and }\\  
f(s_2) \in \GCL(\{s_2\} \cup C) \textnormal{, and } f(s_1) \notin
\GCL(\{s_2\}   \cup C ) \textnormal{, and }\\
f(s_2) \in \GCL(\{s_1\} \cup C')  \textnormal{, and } f(s_1) \notin
\GCL(\{s_1\}  \cup C' ) \textnormal{, and }\\
 f(s_1) \in\GCL(\{s_2\} \cup C') \textnormal{, and } f(s_2) \notin \GCL(\{s_2\}
 \cup C' ).
\end{eqnarray*}
\label{def:const_iboc}
\end{definition}

As stated in Definition~3 each binary molecular code comes 
with a second code
implementing a different mapping. The alternative
code $g$ is determined by $g(s_1) = f(s_2)$ and $g(s_2) =
f(s_1)$. $K = \langle f, g \rangle$ is called {\it code pair}. 
Two simple example networks are shown in Figures~\ref{fig:examplenet}A~and~\ref{fig:examplenet}C. Both networks appear to be very similar in their
structure, but they show different number of codes. While the former network is capable to realize one code pair,
the latter network -- though being smaller -- can realize two code pairs.

\subsection*{Semantic Capacity}
\label{sec:semantic-capacity}
Now we can measure the semantic capacity of a system as the system's
capacity to realize contingent mappings.
Concretely, we count how many different mappings a network can realize.
Note that these codes need not be realized at the same time. Further
note that we count each mapping only once, even if it can be realized
by more than one codemaker.

Because we study only binary codes here and binary codes always come
in pairs, it is reasonable to count the number of different code pairs $CP_N$ of
a network $N$. So, the {\it semantic capacity} is defined here as: 

\begin{equation}
\mathcal{SC}(N) = CP_N.
\end{equation}
The number of code pairs can be high and can grow exponential with
network size, such that we use the logarithm for
comparing different network's semantic capacity. The {\it logarithmic
semantic capacity} is defined as
\begin{equation}
\mathcal{SC}_{log}(N) = \log_2(1+\mathcal{SC}(N)) = \log_2(1+ CP_N).
\end{equation}
In $\mathcal{SC}_{log}$ we apply the transformation $1+x$ to guarantee that
the logarithmized semantic capacity is well defined and its smallest
value is zero, in case the network cannot realize any molecular code.

The mean semantic capacity of a group of $n$ networks $N_i$
($i=1,\dots,n$) is calculated by the arithmetic mean of the respective
measure (linear or log).

In future studies, the semantic capacity could be integrated with measures of the code's
quality, fitness, or cost \cite{Tlusty2008,Tlusty2008b}. E.g., two networks
with the same number of code pairs could be differentiated with
respect to the costs to implement those codes.

\subsection*{Algorithmic Identification}
\label{sec:algor-ident-molec}
The closure-based definition of BMCs (Definition~3)
allows us to develop algorithms for automatic detection of codes in
reaction networks. The algorithm searches for a network pattern,
i.e., a combination of molecular species and reactions, fulfilling the
conditions stated above. 

Definition 3 leads to an algorithm that first calculates all closed
sets and then checks 
combinations of closed sets  for the BMC condition as stated in Definition~3. In
particular for the two elements of the domain, and the two elements of
the codomain the single molecular closed sets, i.e., the
closed sets that are  generated by a single molecular species alone
($G_{CL}(m), m\in \Mol$), are used. There exist at most $|\Mol|$
single molecular closed sets. The closure-based algorithm has a worst-case
running time complexity of ${\cal O}(|\Mol|^4 n_c^2)$ with
$n_c$ as number of all closed sets  contained in the system
(cf. Supplement Text S1).

Alternatively we can analyze a network in terms of pathways. This
makes sense since signs and meanings need always be connected by paths
of reactions.
The running time complexity of the \textit{pathway-based} algorithm depends on
the number of paths the network contains. For the identification of
BMCs all s-t-paths for all pairs of species are identified. Any
combination of four paths is checked for the BMC condition. Since the number of
paths in a network grows enormously with the density of the network we
apply a parameterized algorithm that uses only the $k$-shortest paths \cite{Martins2003}
between every pair of species. The worst case running time then is bounded
by ${\cal O }(|\Mol|^4k^4)$. If $k$ is chosen too small the
algorithm is not able to find all codes in the system, but gives an
approximate measure.  Pseudo-code for both algorithms and subroutines
is given in the supplement (Supplement Text S1).

The different running time complexities suggests a conditional
application of the algorithms. The
pathway-based algorithm can be efficiently applied on networks that have a high number of closed
sets and a low number of paths, while the closure-based algorithm can
be applied in the other case, where the number of paths is high and
the number of closed sets in the network is low.
Interestingly, systems with high semantic capacity tend to have both,
high number of closed sets and many pathways (see below), so that an algorithmic
challenge remains for analyzing such systems more efficiently.

\section*{Results}
\label{sec:results}
In the following we survey different kinds of systems for their
semantic capacity by the application of the algorithms described
above. The analyzed systems are the gene translation chemistry (GC), gene regulatory networks (GRN), phosphorylation
cascades (PC), combustion chemistries (CC), the martian atmosphere
chemistry, and random networks. A summary of the analysis of
the biological systems (GRN, GC, PC) is given in Table
\ref{tab:systems_overview}, while all systems are compared in
Table~\ref{nbsystems}. 
To apply our algorithms we had to
construct the reaction networks for some of these systems first. To
accomplish this we followed a knowledge-based approach for GC, GRN, PC. 

\subsection*{Biological Reaction Networks}

\subsubsection*{Gene Translation}
\label{sec:genetic-code-1}
We will show now that the gene translation chemistry can realize
molecular codes. In particular, this suggests that the genetic code, i.e., the mapping describing
the translation from nucleotide triplets to amino acids, is a molecular code. 
The fact that there is more than one genetic code is known for a
long time \cite{Osawa1992,Jukes1993}. Here, we analyze the 17 already
known genetic codes, as listed at NCBI \cite{Elzanowski2010}. The different
genetic codes cover nuclear and non-nuclear codes of different genera,
e.g., bacterial, archaeal, and plant plastid codes, the vertebrate,
invertebrate, and yeast mitochondrial codes, and the alternative yeast
nuclear code. In particular, we construct a reaction network
containing the codons, the amino acids, and the specific tRNAs, which
are necessary for the translation. For all mappings between DNA
triplets and amino acids occurring in the 17 codes we add a 
reaction in the network of the form $codon + tRNA \rightarrow amino \ acid$.
The obtained reaction network (Supplement Network S2) represents a merge of the 17 genetic codes
and contains 234 molecular species and 85 reactions.

\begin{table}[h]
  \centering
  \caption{{\bf BMCs found in the merge of the 17 known genetic
      codes.} Here the 16 found BMCs are summarized. If applicable
    BMCs are grouped. See supplement Text S3 for the code pairs.}
  \begin{tabular}{llll}
    sign (codons) & meanings (amino acids) & \#BMC & References\\\hline
    CTT, CTG, CTA, CTC & L, T & 6 & \cite{Osawa1992,Clark-Walker1994}\\
    AGG, AGA & G,S,R, Stop & 6 & \cite{Himeno1987,
      Jacobs1988,Batuecas1988,Osawa1989b,Garey1989,Ohama1990,Osawa1992,Hoffmann1992,Durrheim1993,Boore1994,
      Kondow1999,Telford2000,Yokobori2003}\\
    AGG, TCA & S, Stop & 1 & \cite{Batuecas1988,Osawa1989b,Osawa1992, Hoffmann1992, Boore1994, Nedelcu2000}  \\
    AGA, TCA & S, Stop & 1 & \cite{Batuecas1988,Osawa1989b,Osawa1992, Hoffmann1992, Boore1994, Nedelcu2000}  \\
    TTA, TAG & L, Stop & 1 & \cite{Osawa1992,Hayashi-Ishimaru1996,Laforest1997,Nedelcu2000,Elzanowski2010}\\
    TAA, TAG & Q, Stop & 1 & \cite{Osawa1992,Schneider1989,Schneider1991,Liang1993,Keeling1996}\\\hline
  \end{tabular}

  {\scriptsize References -  Articles reporting the respective alternatives in the genetic
  code that are part of a BMC in this analysis.}
  \label{tab:gc_merge_results}
\end{table}

The algorithmic analysis of this network identified 16 binary
molecular codes (Supplement Text S3), i.e., a semantic capacity of $SC_{log}=4.09$. The binary codes can partly be assigned to larger molecular
codes. CTT,CTG,CTA, and CTC can be mapped on
leucin (L) and threonin (T) and give rise to six of the found BMCs. The second
group involves the mapping between AGG,AGA and glycin (G), serine (S),
arginine (R) and the translation stop. This code can also be decomposed
into six BMCs. There does exist four more BMCs that involve the codons
TCA, TTA, TAG and TAA and the amino acids leucine (L), glutamine (Q) and the
stop signal. Table \ref{tab:gc_merge_results} summarizes the BMCs
found.  The existence of alternative mappings in the genetic 
translation system suggests that the genetic code qualifies as
a molecular code.

We may model the genetic code now by including all potential mappings
between codons and amino acids, i.e., the model includes all possible
tRNAs such that any codon could be read for any amino acid (Supplement
Network S3). In such a
system the number of binary molecular codes can easily be
calculated. Each pair of codons forms a code pair with each pair or amino
acids. Since there exist $64 \choose 2$ pairs of triplets and $20
\choose 2$ pairs of amino acids the number of BMCs in this potential
set up is 
\begin{equation}
{64 \choose 2} \cdot {20 \choose 2} = 383,040.
\label{eqn:gc-full}
\end{equation}
The logarithmic semantic capacity is $\approx 18.55$.
\begin{figure*}[!ht]
  \centering
  \includegraphics[width=1\textwidth]{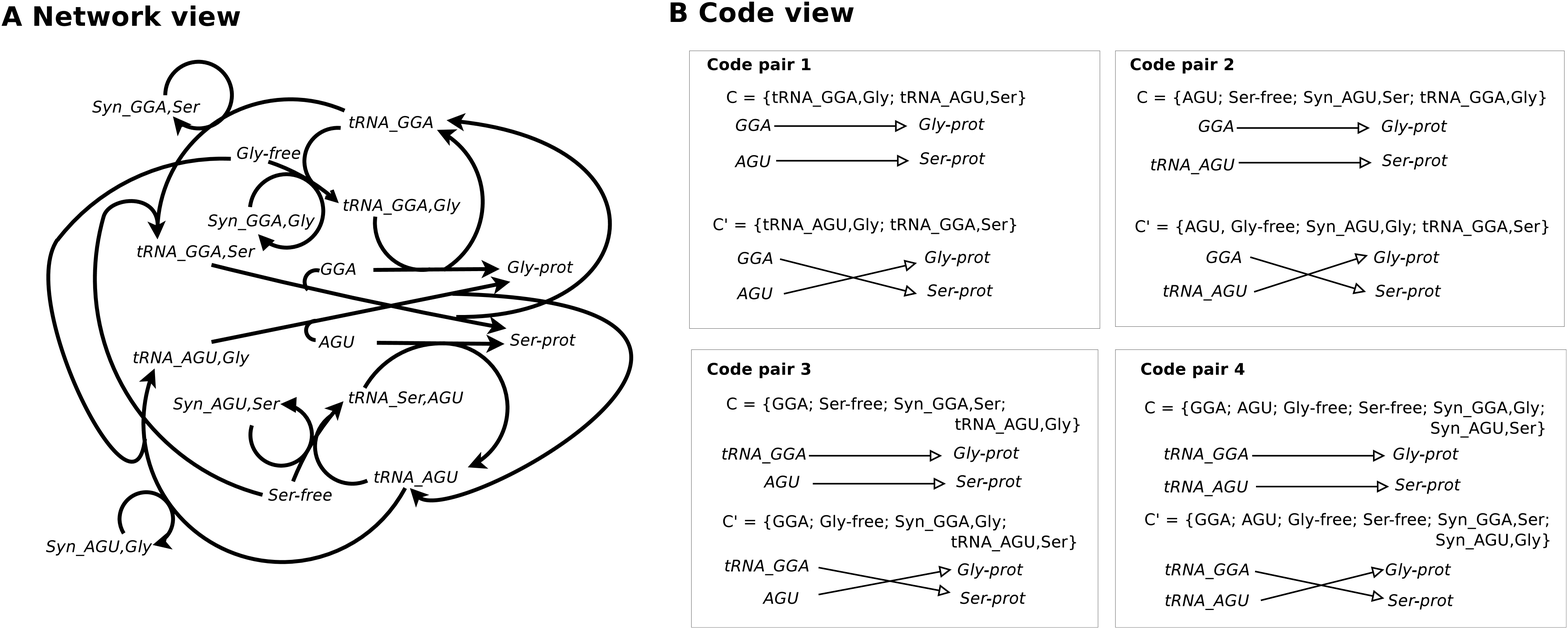}
  \caption{{\bf Subnetwork of the full gene translation network model
      with synthetases ($N_{GC}$).} The
    network (panel A) shows a subnetwork of the gene translation
    network model containing the translation, and loading reactions
    for two selected codons (GGA,AGU) and amino acids (Gly,Ser). The
  semantic analysis shows that four code pairs can be implemented by
  this network (panel B).}
  \label{fig:gctoy}
\end{figure*}

In the following, we refine the network model by constructing a reaction
network containing all possible mappings between the 64 codons and 20 amino
acids like described above. Additionally we model the loading step of the tRNAs by
inserting the respective amino-acyl-tRNA-synthetases (aaRS)
(cf. Figure \ref{fig:gctoy}, Supplement Network S5).
The reaction network $N_{GC} =  \langle \Mol_{GC},\Rea_{GC}
\rangle$ describes the core molecular mechanism realizing the standard genetic
code and all alternative codes. The set of molecular species
$\Mol_{GC}$ of the network contains all DNA strings of length
three (Table \ref{tab:gc}, Eq. 2), representing the codons. It
contains the twenty proteinogenic amino acids in their free form
(Table \ref{tab:gc}, Eq. 3) and the twenty amino acids bound in a
protein (Table \ref{tab:gc}, Eq. 4).  To describe the system properly we
also need to insert species for all possible tRNAs in their unloaded (Table \ref{tab:gc}, Eq. 5) and loaded form
(Table \ref{tab:gc}, Eq. 6). In the unloaded form we represent specificity to codon $n$
with $n$ as subscript, in the loaded form we represent specificity to
$n$ with a subscript $n$ and the loaded amino acid $a$ with a
subscript index $a$, i.e., a tRNA that is loaded with $Ser$ and has
specificity to $AGU$ is denoted as $tRNA_{AGU,Ser}$. The network also
contains all possible aaRS (Table
\ref{tab:gc}, Eq. 7), $Syn_{n,a}$, such that the system is able to
load all amino acids to all tRNAs. The specificity of the aaRS to
certain combinations of codons and amino acids is represented by two
subscript indices $n,a$, with $n$ representing the respective codon
and $a$ representing a certain amino acid. The set $\Rea_{GC}$
contains all reactions loading the amino acids onto the tRNAs (Table
\ref{tab:gc}, Eq. 8) and all reactions inserting an amino acid in 
the peptide sequence (Table \ref{tab:gc}, Eq. 9). Figure \ref{fig:gctoy}A displays a subnetwork with
two codons (GGA,AGU), two amino acids (Gly,Ser) and the respective other elements of the
network (tRNA and synthethases).

\begin{table*}[!ht]
  \small
  \caption{{\bf Reaction network formulation of a gene translation
      system with synthetases}}
  \begin{tabular}{lp{10cm}p{7cm}}
    Eq. & Definition & Description \\\hline
    1 & $\Mol_{GC} = Codons \cup AA^{free} \cup AA^{prot} \cup aaRS \cup tRNA^{free}
    \cup tRNA^{loaded}$ & Definition of the molecular species in the
    network \\
    2 & $Codons=\{A,C,G,T\}^3 = \{AAA, AAC, \dots,
    TTT \}$ & Set representing the 64 codons of the genetic code\\ 3 &
    $AA^{free} = \{Ala^{free}, Arg^{free}, Asp^{free}, \dots,
    Try^{free}\}$ & Amino acids that are not used in a protein\\ 4 &
    $AA^{prot} = \{Ala^{prot}, Arg^{prot}, Asp^{prot}, \dots,
    Try^{prot}\}$ & Amino acids that have been used in a protein
    during gene translation\\ 5 & $tRNA^{free} = \{tRNA_{n} | n \in Codons\}$ & Unloaded tRNAs specific for codon $n$\\ 6 &
    $tRNA^{loaded} = \{tRNA_{n,a} | n \in Codons, a \in
    AA_{free}\}$ & tRNAs specific for codon $n$ that have been loaded
    with amino acid $a$\\ 7 & $aaRS = \{Syn_{n,a} | n \in
    Codons, a \in AA_{free}\}$ & Amino-acyl-tRNA-synthetases that are
    specific for amino acid $a$ and codon $n$ \\ 8 & $\Rea_{GC} =
    \{tRNA_{n} + a + Syn_{n,a} \rightarrow tRNA_{a,n} + Syn_{n,a} \ |
    \ n \in Codons, a \in AA^{free} \} \cup $ & Loading of the
    tRNA by suitable synthetasis \\ 9 & $\{n + tRNA_{a,n} \rightarrow
    n + tRNA_{n} + a \ | \ n \in Codons, a \in AA^{prot} \}$ &
    Translation step, i.e., the incorporation of an amino acid into a
    growing protein\\

  \end{tabular}

  \label{tab:gc}
\end{table*}

The analysis of this extended network ($N_{GC}$) describing all potential genetic
codes with $64$ codons and $20$ amino acids results in $1,532,160$
binary code pairs, i.e., $SC_{log}(N_{GC})\approx 20.55$. This is a different result than for the less detailed
model, as calculated by Eq. (\ref{eqn:gc-full}). The extension of the model by aaRS, unloaded
tRNAs, and unloaded amino acids increases the semantic capacity. A
closer look to the resulting codes shows that not only the codons can
be signs, but also the unloaded tRNAs ($tRNA^{free}$) can function as signs. These additional signs increase
the number of code pairs. The ``new'' codes differ structurally in
their codemakers. While, classically, the codons are mapped to the set of amino
acids  ($AA^{prot}$) using the loaded tRNAs ($tRNA^{loaded}$) as
codemakers, the new signs, i.e., unloaded tRNAs, are mapped to the set of amino acids by 
using a codemaker that consists of the free amino acid loaded to the
free tRNA, the synthetase performing the loading step, and the codon
that needs to be recognized by the tRNA.  
The number of code pairs in this system can be calculated by 
\begin{equation}
\left({n_s \choose 2}-\frac{n_s}{2}\right)\cdot{n_m \choose 2},
\label{gc-ex-full}
\end{equation}
with $n_s$ as number of signs and $n_m$ as number of meanings. For the
full gene translation system the number of signs is
$n_s=|Codons|+|tRNAs^{free}|$ and $n_m=|AA^{prot}|$. Since there is
always one pair of one tRNA and codon belonging together, which
therefore can not be combined in an BMC, we have to subtract the
number of such pairs $n_s/2$ from the amount of all combinations. 

\begin{table*}[!b]
\centering
\footnotesize
\caption{{\bf Code pairs realized by the subsystem of the gene
    translation network with synthetases shown in Figure \ref{fig:gctoy}.}}
\begin{tabular}{lllll}
 & Signs  & Meanings \\\hline
Code pair 1 & 
$\{GGA, AGU\}$ & $\{Gly^{prot}, Ser^{prot}\}$ \\
Code pair 2 & 
$\{GGA, tRNA_{AGU}\}$ & $\{Gly^{prot}, Ser^{prot}\}$ \\
Code pair 3 & 
$\{AGU, tRNA_{GGA}\}$ & $\{Gly^{prot}, Ser^{prot}\}$ \\
Code pair 4 & 
$\{tRNA_{GGA}, tRNA_{AGU}\}$ & $\{Gly^{prot}, Ser^{prot}\}$\\
\hline
\end{tabular}
\label{tab:gc_toysystemA}
\end{table*}

\begin{table*}[!b]
\centering
\footnotesize
\caption{{\bf Codemakers of the code pairs shown in Table \ref{tab:gc_toysystemA}.}}
\begin{tabular}{lp{6.5cm}p{6.5cm}}
Code pair & Codemaker & alternative Codemaker \\\hline
1 & $\{tRNA_{GGA,Gly}, tRNA_{AGU,Ser}\}$ & $\{tRNA_{AGU,Gly}, tRNA_{GGA,Ser}\}$ \\ 
 2 & $\{AGU,Ser^{free}, Syn_{AGU,Ser}, tRNA_{GGA,Gly}\}$ &
$\{AGU,Gly^{free}, Syn_{AGU,Gly}, tRNA_{GGA,Ser}\}$ \\ 
 3 & $\{GGA,Ser^{free}, Syn_{GGA,Ser}, tRNA_{AGU,Gly}\}$ &
$\{GGA,Gly^{free}, Syn_{GGA,Gly}, tRNA_{AGU,Ser}\}$ \\ 
 4 & $\{GGA, AGU, Gly^{free}, Ser^{free}, Syn_{GGA,Gly},$ $Syn_{AGU,Ser}\}$ &
$\{GGA, AGU, Gly^{free}, Ser^{free}, Syn_{GGA,Ser},$ $Syn_{AGU,Gly}\}$\\
\hline
\end{tabular}

\label{tab:gc_toysystemB}
\end{table*}

\subsubsection*{Gene Regulatory Networks}
\label{sec:gene-regul-netw}

A gene regulatory network (GRN) is a graph representing the regulation
of gene expression of certain genes by the expression of other genes.
A node in a GRN stands for a complex
process. It represents the gene, the promoter and binding region of
that gene, the binding of the transcription factor (TF) plus cofactors and
the production of a product by the recruitment of the gene expression
machinery.  

We will show here that the GRN of a cell is also a highly
semantic system. Different sources of signals are integrated into a
decision which gene is transcribed and which is repressed. Each of
these signals has another meaning which emerges out of the contingency
of the system. The system is contingent, because a mapping between
signal and gene product can be altered by the exchange of a promoter
region of a gene (or vice versa). This may happen enzymatically by the
application of nucleases and ligases or by mutations.

To identify the semantic capacity we \RED{describe} a gene
regulatory network as a reaction network ${\cal N}_{GRN} = \langle
\Mol_{GRN},\Rea_{GRN} \rangle$.
$\Mol_{GRN}$ contains $n$ transcription factors
$\TF_i$, $m$ products $P_i$, and genes $G_{ij}$.
Each gene $G_{ij}$ represents a combination
of a promoter site $i$ and a coding region $j$, where the promoter
site $i$ is specific to $TF_i$ and the coding region $j$ produces $P_j$.
For our model we assume that there exist as many promoter sites and coding regions as transcription
factors and products, respectively, such that any gene is possible.
The set of all species $\Mol_{GRN}$ then is 
\begin{eqnarray*}
\Mol_{GRN} = \{\TF_1,\TF_2,\dots, \TF_i,\dots, \TF_n,P_1,P_2,\dots,\\P_j,\dots,P_m,G_{11},G_{12},\dots,G_{ij},\dots,G_{nm}\}.
\end{eqnarray*}
Assuming that a transcription factor only binds one promoter and that a
promoter is bound only by one by one transcription factor the
expression of a gene $i,j$ is given by 

\begin{eqnarray*}
\Rea_{GRN} = \left\{ \TF_i + G_{ij}  \rightarrow \TF_i + G_{ij} +   P_j \right\},  i =
1,2,\dots,n,\\ j = 1,2,\dots,m.
\end{eqnarray*}
\begin{figure*}[!ht]
  \centering
  \includegraphics[width=0.8\textwidth]{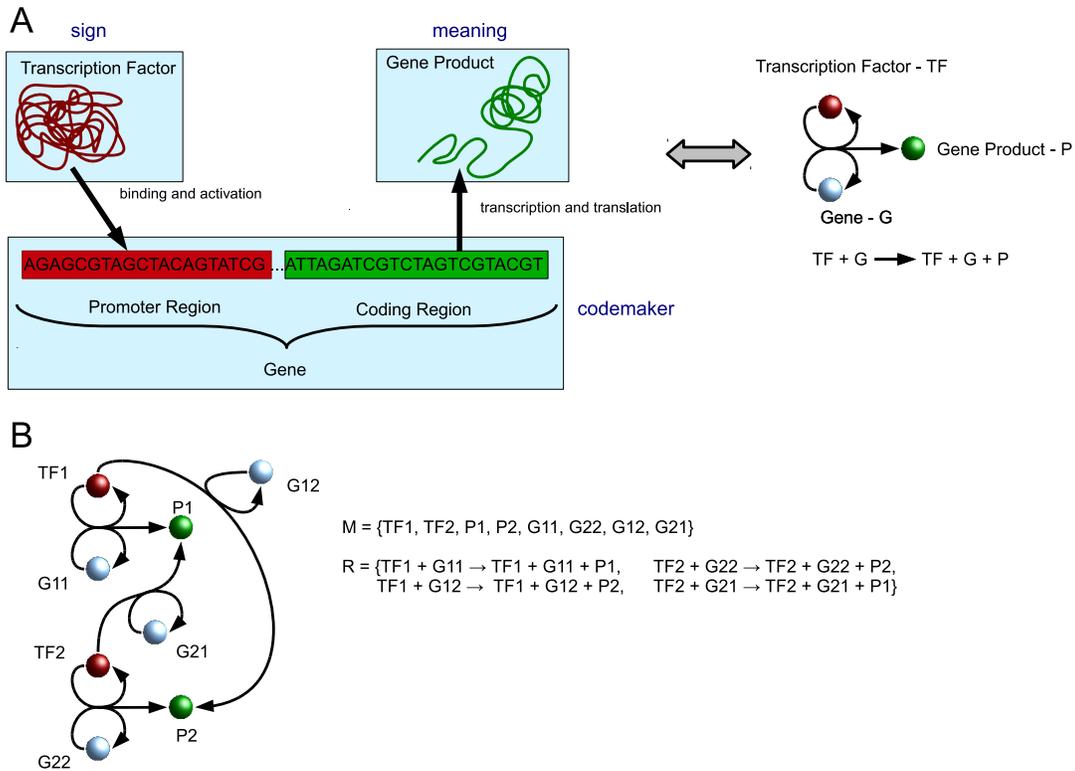}
  \caption{{\bf Gene regulatory network model.} Panel~A: (left) Simple model
    of the expression of a gene, (right) reaction network formulation
    of the same process.
Boxes in Panel~A indicate the semantic interpretation, i.e., the
transcription factors are the signs, the products are the meanings, and the
DNA is the codemaker.
Panel~B: reaction network constructed according to the formalization
of gene regulation shown in (A).
}
  \label{fig:grnmodel}
\end{figure*}

Figure \ref{fig:grnmodel} illustrates the network definition.
We here do not present a generic model to describe all possible gene regulatory
networks, but a model that covers the main properties of regulation important
for this study.
The analysis of this system shows that the reaction network can
implement molecular codes only in one way, i.e., with the transcription factors
as signs and the set of products as meanings.
The set of genes, i.e., the combination of promoter and coding region, forms the codemaker, because it allows for the
contingent implementation of mappings between signs and meanings. 
Thus, in contrast to the model of the gene translation chemistry described above, the DNA is not the sign,
but functions as the codemaker, i.e., it carries the contingency of
the system. This shows that a code based analysis can only be done
with regards to systems and not to single molecular species.

\subsubsection*{Signaling Networks: Phosphorylation Cascades}
Cells maintain signaling systems of different kinds
for signal transmission and integration \cite{Krauss2008}.  
The most prominent signaling systems rely on reversible
phosphorylation of amino acids side-chains for regulation of signaling
protein activity. The direct involvement of such systems in signaling
suggest that they may be also semantic systems. If so they should be
able to realize molecular codes. 
We have studied phosphorylation cascades, like the MAP kinase regulatory network, as a
typical instance of an intra-cellular signaling system. 
These systems demonstrate the limitation of our closure-based
approach. The static approaches described above are not sufficient to
detect the codes in a phosphorylation cascade.  A more refined
approach is necessary that distinguishes between concentrations. It
can be derived from our definitions here in a straight forward
way. 

\begin{figure*}[!ht]
\centering
\includegraphics[width=0.7\textwidth]{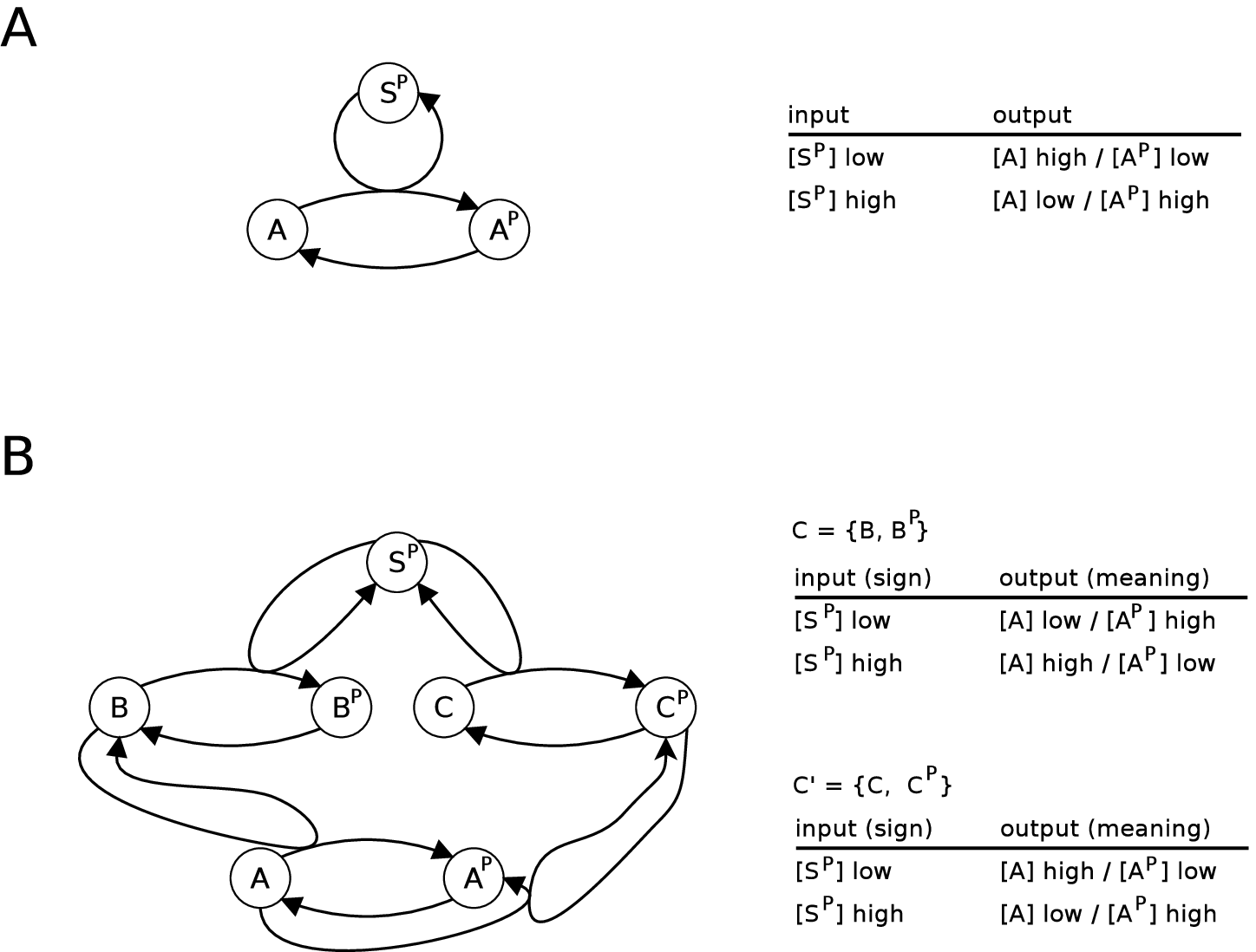}
\caption{
{\bf Reaction networks describing phosphorylation motifs.} 
Molecular species in these networks represent kinases or phosphatases
that may be activated or inactivated. Activated and  non-activated forms
of a kinase/phosphatase are modelled as different species (e.g., species A/A$^P$). 
Panel A: (left) Reaction network of a simple phosphorylation motif,
which can realize a molecular mapping (right), but not a molecular code;
panel B: (left) more complex reaction network that can realize a
molecular code. The molecular code is not only specified by the
species, but also by their concentrations.
}
\label{fig:phoscasc}
\end{figure*}

Applying this more refined approach, i.e., taking concentrations
into account, we can see that the activation of a kinase by 
phosphorylation can generate a molecular mapping, but this mapping is not
necessarily a molecular code (Figure~\ref{fig:phoscasc}A). A two-step
cascade is able to implement a molecular code (Figure
\ref{fig:phoscasc}B).

The simple one-step phosphorylation model (Figure \ref{fig:phoscasc}A)
contains two kinases; an initial kinase ($S$) which is able to
phosphorylate the target kinase $(S^P \ + \ A \rightarrow A^P$). We
also model the dephosphorylation ($A^P \rightarrow A$). For sake of
simplicity we do not model the phosphatases, and the phosphate related
molecular  species (e.g., ATP, ADP, P) involved in the process, but assume a buffered
concentration. In the simple, one step, model we may observe a
molecular mapping between $S^P$ and the two states of kinase $A$. If
$S^P$ has a low concentration the system is  in a state where the
unphosphorylated state $A$ has a high concentration and the
phosphorylated state $A^P$ has a low concentration. According to the
definition of molecular code given above the system should be able to
change the mapping, i.e., be contingent, by the application of a
different molecular context to realize a code. Here, no alternative
mapping between $S$ and $A$ can be realized, such that the
system is not able to realize a molecular code.

If we consider a different system where two kinases are inbetween $S$ and $A$,
we obtain a two-step phosphorylation cascade. $S^P$ now phosphorylates
the inserted species, while these have an effect on $A$. Now the system has
the possibility to  ``choose'' between two alternative systems, i.e.,
the inserted species may be ``active'' in the unphosphorylated state
($B$), or in the phosphorylated state ($C$). There exist several mappings in
such a system, e.g., between $S$ and $B$, $S$ and $C$, and $S$ and
$A$. The former two mappings behave like the simple model (see
above). The mapping between $S$ and $A$ is a molecular code, because
the molecular context of the system can be changed, such that the
alternative system behavior is generated (see Figure
\ref{fig:phoscasc}B (right)). The codemaker of the code between $S$
and $A$ is either the $B$-system, or the $C$-system.

\subsection*{Random reaction networks as null model for code identification}
\label{sec:artif-chem}

To check whether the motif describing a BMC can be generated by
chance we analyzed random reaction networks of different sizes and
densities for their semantic capacity. The networks have been
generated by random insertion of reaction rules in the empty
network. Each random reaction rule is bimolecular, i.e., contains two
reactants, and one product. 

\begin{figure*}[!ht]
\centering
\includegraphics[width=0.8\textwidth]{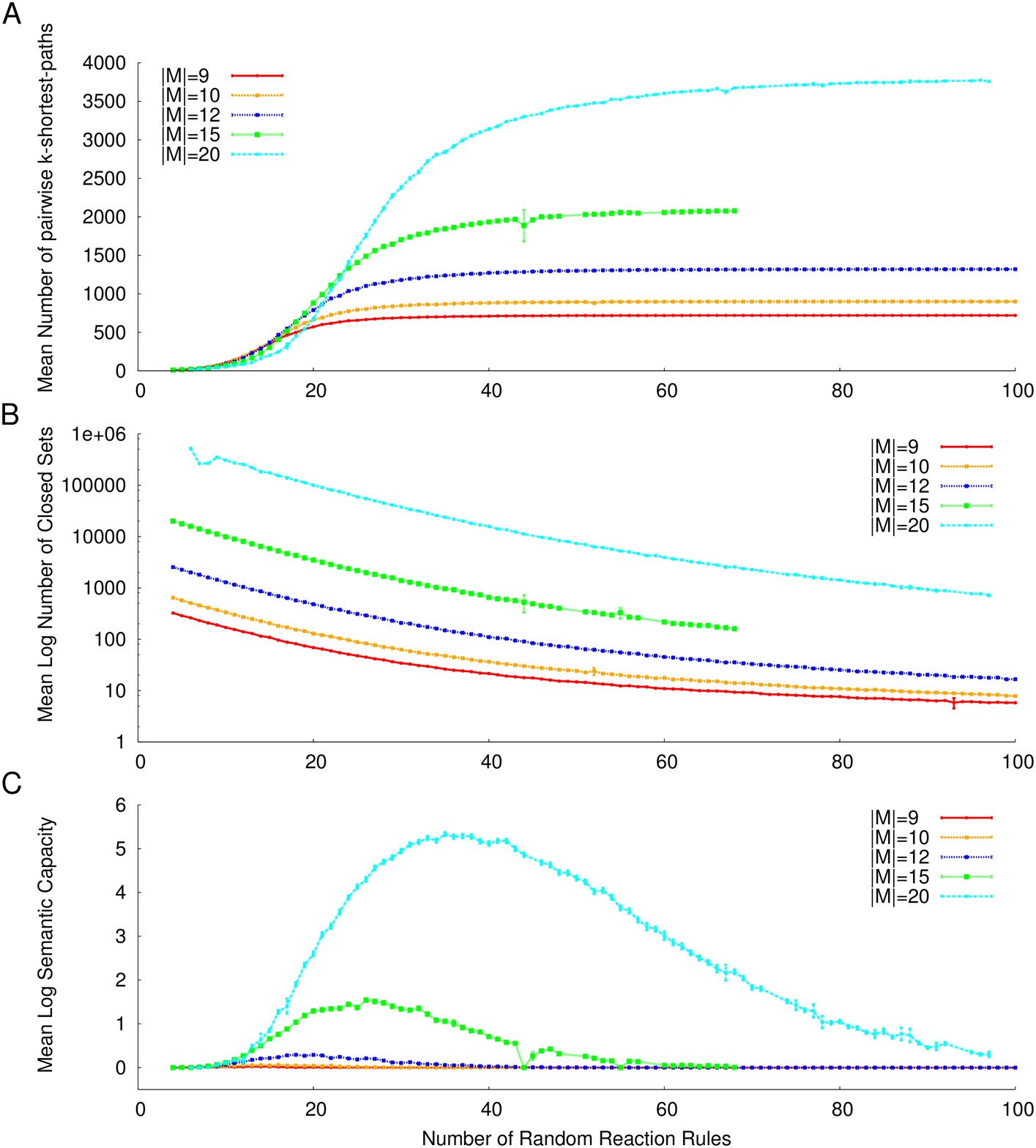}
\caption{
{\bf Structural properties of random reaction networks.} 
Panel A: Number of paths of the respective combination of species and reactions.
Panel B: Number of closed sets.
Panel C: Semantic capacity.
Panels A,B and C show three important network parameters for five
different network sizes and various numbers of reaction rules. Each
data point represents the average about random replicates. Error bars
indicate the standard error of the mean. Panel A shows the
average number of paths in the network. Since we applied the
path algorithm which only uses the k-shortest paths between each pair
of molecular species the curve shows a sigmoidal behavior, which is
saturated at the value ${|\mathcal{M}| \cdot( |\mathcal{M}|-1)} \cdot
k$, with $k=10$ .
Panel B shows the average number of closed sets. With growing density
the number of closed sets decreases. Panel C shows the distributions of the average number of
code pairs (log measure with basis 2) in random networks of different
sizes. The semantic capacity shows an unimodal distribution, which
correlates with the other two shown network parameters. If the number
of paths is too low no mappings can be implemented because of the
missing links. If the number of closed sets is too low no unique
mappings can be implemented.
}
\label{fig:rrn}
\end{figure*}

\begin{figure*}[!ht]
\centering
\includegraphics[width=0.5\textwidth, angle=270]{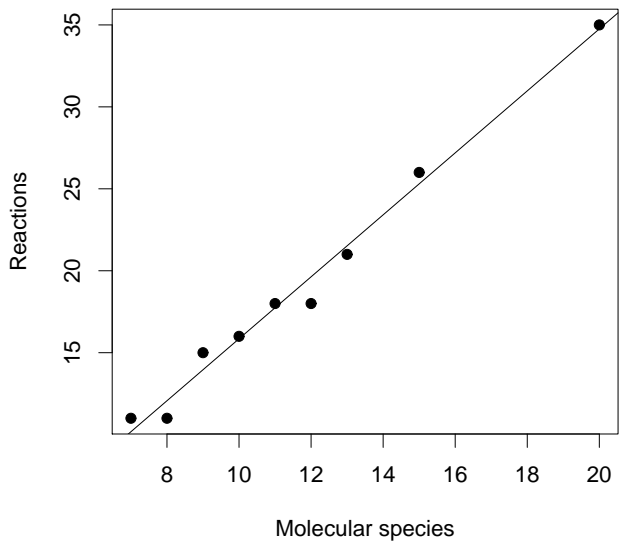}
\caption{
{\bf Scatter plot of the positions (Molecular Species, Reactions) of
  the maximal semantic capacity of the unimodal distributions for the random
  networks analyzed in this study.}  The linear regression results in the function: $Reactions = -3.06 +
1.89 \cdot Molecular\ species$. 
}
\label{fig:rrn_max}
\end{figure*}

The analysis showed that the binary code motif can be generated in
random networks (see Figure \ref{fig:rrn}), i.e.,
contingent mappings can be generated randomly.
For a fixed network size and varying densities the average semantic
capacity shows a unimodal behavior, which suggests that there exist
an optimal range of densities for each network size, leading to maximal
semantic capacity. This optimal range shifts to higher densities
with increasing size of the network (see Figure \ref{fig:rrn_max}).  The optimal interval is bounded on the
left (lower densities) by the low complexity of the network, there are
not enough reactions to promote the insertion of molecular by
chance. On higher densities the network is strongly connected, such
that it is harder to obtained closed sets, and
therefore it is also harder to implement codes by chance.
The optimal interval coincides with two important network properties,
i.e., the number of paths, and the number of closed sets. With
increasing network density the number of paths grows, while the number of closed sets
decreases. High semantic capacity can be found in networks with a  high number of
pathways and at the same time a high number of closed sets.

\subsection*{Non-Biological Reaction Networks}
\subsubsection*{Combustion Chemistries}
\label{sec:non-biol-syst}
We analyzed a number of chemical systems, i.e., combustion chemistries
of hydrogen \cite{Conaire2004}, methane \cite{MethaneOxidation2001},
ethanol \cite{Marinov1999}, dimethyl ether \cite{Kaiser2000}.
\RED{
The original combustion chemistry data (provided in CHEMKIN format \cite{Keeling1996}) have been processed
to obtain the reaction networks <describing the respective
chemistry. The chemistries are intended to describe all significant
processes that  can occur in
the combustion, i.e., burning, of the respective molecule. In the
CHEMKIN files most of the reactions are described as equilibrium
reaction with additional thermodynamic parameter. Taking these as
basis we obtain reaction networks (see definition above) containing
the directed reactions depending on the thermodynamic parameters. 
The obtained reaction networks (Supplement S7) vary in their size (10 - 79 molecular species) and
density (38 - 752 reactions). 
We found that none of the chemistries is able to realize molecular
codes. 

We also analyzed the atmosphere chemistry of Mars \cite{Nair1994} to
check whether other kinds of non-biological chemistries may contain
codes. The atmosphere chemistry of mars contains 32 molecular species, 104
reactions and 5512 closed sets. In particular the network describes
the reaction happening on the day side of mars. Therefore, light ($h\nu$)
is modelled explicitly and there exist an inflow reaction for light.
The Martian chemistry also is not able to realize molecular codes. 

Here, we compare the obtained results with random reaction networks of
 same size and density. 
The original hydrogen chemistry could not realize molecular codes. This
may be due to the small number of closed sets compared to the number of
paths, such that the molecular species are ``too connected'' and the network is less
structured. In random networks of same size and density no molecular code can be identified. The
estimated number of closed sets and paths, although differing from the
from the original chemistry, are also marking that the networks are
not in the optimal interval  (compare section \ref{sec:artif-chem}).

In the methane combustion chemistry we see that there exist far more
paths than closed sets, such that the network is ``unstructured''. The
according null-model, here, also contains a high number of paths,
but also a higher number of closed sets. The algorithmic analysis showed
that some null-model networks can realize BMCs, such that the
average semantic capacity is $SC_{log} = 1.04$.  Nevertheless we
consider this also as a very low semantic capacity compared
with, e.g., the gene translation chemistry.
For the other two combustion chemistries (ETH, DME) and the Martian atmosphere
chemistry (MARS) the analysis of the random networks is not feasible
with our current algorithms, due to the large number of paths and closed sets in these networks.
}

\begin{table*}[!h]
  \footnotesize
  \caption{{\bf Comparison of combustion chemistries and random
      networks (null model).}}
  \begin{tabular}{lllllllll}
    \multicolumn{6}{l}{Combustion chemistry properties}   &
    \multicolumn{3}{l}{Null model estimate} \\
&$|\Mol|$&$|\Rea|$&\#closed sets&\#paths&$SC_{log}$&est. \#closed sets (s.e.)&est. \#paths (s.e.)&est. $SC_{log}$ (s.e.)\\\hline
HYD&10&38&16&$7.69 \cdot 10^4$&0&39.8 (0.53)&878.2 (1.27)&0 (0.0)\\
MET&37&340&4,136&$>10^{6^\star}$&0&6,521.83 (353.63)& $>10^{6^\star}$ & 1.09 (0.15)\\
ETH&57&752&5,136&$>10^{6^\star}$ &0&$>10^{5^\star}$&$>10^{6^\star}$ &n.a.\\
DME&79&708&8&$>10^{6^\star}$&0&n.a.&$>10^{6^\star}$ &n.a.\\\hline
  \end{tabular}
  n.a. - not available, s.e. - standard error of the mean,  $^\star$estimated
  by performing runs on several networks (or growing values of k
  regarding \#paths) where not all runs completed due to computational
  complexity, such that the maximal found value gives the estimate.
  \label{tab:comb-rand}
\end{table*}

\subsubsection*{Artificial Chemistry NTOP}
\label{sec:ntop}
Recall that with increasing density random networks have a vanishing semantic
capacity. In the following we show that even a dense network can have a 
relatively high semantic capacity.
For this purpose we analyze an artificial chemistry  
with 16-species introduced by Banzhaf~\citet{ac:Ban93a} called NTOP. 
For each species there is a 4-bit binary representation and the
reaction rules are derived with respect to this representation, which
is referred to as a structure-to-function mapping 
(see~\citep{ac:Ban93a} for details and Supplement S8 for the full
network model).

\begin{figure}[htb]
\centering
\includegraphics[width=0.25\textwidth, angle=270]{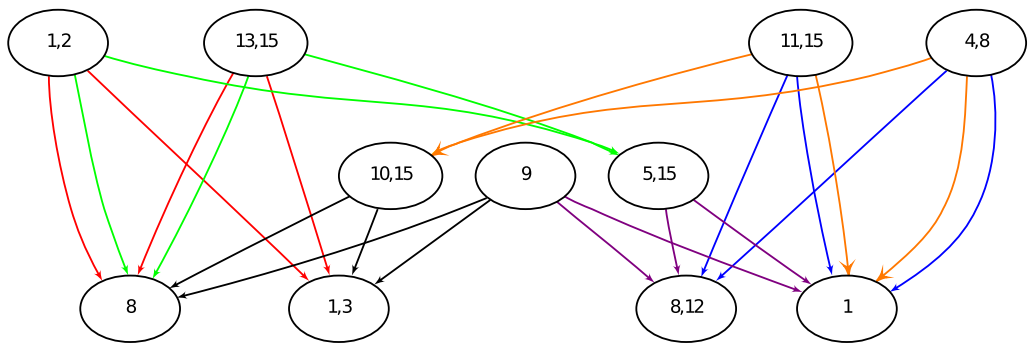}
\caption{{\bf Code pairs found in the artificial chemistry NTOP.} An edge
  connects a sign  with a meaning if they occur in the 
  same code (indicated by colors). Species belonging to codemakers have been
  omitted. Note that a sign can be used in different codes
  and that a meaning might be used as sign in another code
  (e.g., 10 in  \{10,15\}). A vertex represents the closed set generated by
  the respective species alone.}
\label{fig:ntop_codeoverview}
\end{figure}

The algorithmic analysis results in six code pairs (for
an overview see Figure \ref{fig:ntop_codeoverview} and Supplement Text S9). One of the code
pairs with its respective codemaker is shown in Table \ref{tab:ntopiboc}. 
Figure \ref{fig:ntop_codeoverview} illustrates two properties of
molecular codes, (1) a meaning can take the role of a sign in another
code, and (2) molecular species can function as signs (or meanings) in
different codes, i.e., they keep their role in different contexts. Both
properties reflect the context dependency of codes, i.e., the molecular
species constituting the code depend on the molecular context, the codemaker. 

\begin{table}[htb]
\centering
\footnotesize
\caption{{\bf Binary molecular code from the NTOP chemistry.} Species are
  indicated by index. The first line indicates the respective
  species. The second line contains the closed sets generated by the
  species alone and the closed sets that form the codemaker.}

\begin{tabular}{ccccccc}
 & sign 1 & sign 2  & meaning 1 & meaning 2 & codemaker & alternative codemaker\\\hline
species & 2 & 13 & 3 & 8 & - & - \\
closed sets & \{1,2\} & \{13,15\} & \{1,3\} & \{8\} & \{0,5,10,15\} &
\{0,1,9\} \\\hline
\end{tabular}
\label{tab:ntopiboc}
\end{table}

To test the robustness of the network's ability to realize 
molecular codes, we randomized it by replacing 1, 2, 5, 10, 15,
200, and 1000 reaction rules randomly. The number of educts and
products for each individual reaction is kept constant, only the
molecular species are replaced.
Increased randomization result in a clear decline of the average semantic capacity. 
Nevertheless in some cases the randomized network is capable to
implement more code pairs. The average trend, i.e., loss of code
pairs, can be explained by referring to random reaction networks.
Random reaction networks with the same number of species and reactions
as NTOP also have a very low semantic capacity ($SC_{log} = 0$). 
Thus the randomization of the NTOP chemistry drives the
system towards the mean semantic capacity of random networks. 

\begin{figure}[htb]
\begin{center}
\includegraphics[width=6cm, angle=270]{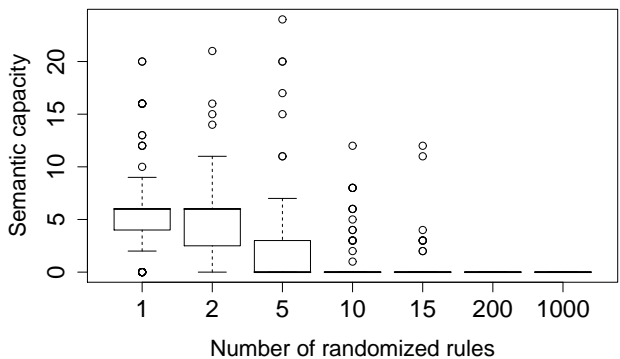}
\caption{{\bf Randomization of NTOP.} Boxplot showing the $SC$-distribution
  of the randomized NTOP chemistry. We can observe that on average
   randomization destroys the semantic capacity
  (non-logarithmized) of the network. The boxplots show the
  distribution of the number of code pairs after randomization of
  NTOP ($n=100$)}
\end{center}
\label{fig:randomized-ntop}
\end{figure}

\section*{Conclusion}
We introduced a formal criterion for identifying molecular codes in
reaction networks and a measure of the semantic capacity of a network,
as the number of different code pairs the network can realize.
Our notion of contingency, defined as the ability of systems to choose between different
mappings, extends the notion of ``independence'' used by Barbieri.

Applying the new concepts to different networks, our basic finding is that the semantic capacity of biological
networks tends to be higher than that of the studied non-biological networks.
Thus, an important step during the transition from non-life to life
must have been the utilization of a chemistry that allows to implement
molecular codes. In our opinion it is an open issue how that
first coding chemistry has looked like.
But we have now a criterion that can guide us in what we have to look
for.

Moreover we can now precisely formulate another hypothesis, namely, that during
the course of evolution the
semantic capacity of the chemistry employed by the biological systems
has a tendency to increase, though not necessarily monotonously. 
One candidate mechanism is the invention and improvement of
compositional adaptors, like proteins with exchangeable domains
\cite{Bornberg2010} or
genes including their promoter- and coding-regions \cite{Barbieri2008}.
Note that also 
the appearance and evolution of neurons and cognitive systems is in
line with the hypothesis of increasing semantic capacity.

\begin{table*}[!t]
\footnotesize
\caption{{\bf Overview of semiotic interpretation of the biological systems surveyed in this paper.}}
\begin{tabular}{lp{5cm}p{5cm}p{5cm}}
role & gene regulatory codes & genetic codes & phosphorylation cascade codes \\\hline
signs & transcription factor & codon or unloaded tRNA & high
concentration of a kinase or phosphatase \\
meanings & gene product & amino acid & high/low concentration of a
target molecule \\
codemaker & DNA with promoter and coding region & loaded tRNA or
mixture of loaded tRNA, aaRS, and codons & a kinase or phosphorylases\\\hline
\end{tabular}
\label{tab:systems_overview}
\end{table*}

The analysis of a formalization of the genetic code showed that not
only the codons are signs for an amino acid, but also tRNAs could be
signs. The bio-molecular and evolutionary interpretation of this fact
should be left for future studies. Furthermore, we have shown that
DNA not only can function as a sign but  also as a codemaker,
as the modeling of GRNs revealed.
The mechanisms in gene regulatory systems and the observation
that such systems are highly flexible (i.e. the mapping between TFs
and products can easily be changed) leads to the conclusion that the
chemistry of GRNs possesses also a high semantic capacity.

The analysis of random networks of different sizes and densities results
in a better understanding of the basal rate of code occurrence. 
We can observe that the distribution of BMCs is unimodal.
Random networks with high semantic capacity show at the same time a
high number of closures (which decreases with increasing network
density) and high number of pathways (which decreases with decreasing
network density). The analysis of an artificial chemistry showed that
also in dense networks the semantic capacity can be high. We hypothesize
that this was caused by the structure-to-function mapping applied
in the artificial chemistry.

Future work includes the formal integration of information theory and
the integration of pragmatics. Furthermore we can extend this static
algebraic approach to a continuous and dynamics approach.

When we address the semantic aspects of biological information,
terminology becomes a hot topic of discussion. Although we appreciate this
discussion from a philosophical perspective, we believe a pragmatic
focus is necessary to obtain a stronger impact in life
sciences. This pragmatic track of the study of meaning for biological
information requires at least three ingredients: (1) (semi-)formal
definitions, (2) algorithm, tools, and predictions, and (3) links to
experimental data (i.e., the physical world).
These three ingredients obviously interact with each other and should
thus be studied together.

\section*{Supporting Information}
\paragraph{Text  - S1} \texttt{s1\_pseudocode.pdf}\\
Pseudocode of all algorithms and subroutines used in the analysis of the data for this paper.

\paragraph{Network - S2} \texttt{s2\_gc\_merge.zip}\\
Network file in rea- and sbml-format containing the description the
merge of the 17 genetic codes listed at NCBI.

\paragraph{Text  - S3} \texttt{s3\_gc\_merge\_codes.pdf}\\
Text file containing the result of the computational analysis of the
network from S2.

\paragraph{Network - S4} \texttt{s4\_gcfull\_64\_20.zip}\\
Network file in rea- and sbml-format containing the description of the
full gene translation chemistry (without synthetases).

\paragraph{Network - S5} \texttt{s5\_gcsynth\_Gly\_Ser.zip}\\
Network file in rea- and sbml-format containing the description of a
subnetwork (Gly,Ser) of the full gene translation chemistry (with
synthetases).

\paragraph{Text  - S6} \texttt{s6\_gcsynth\_Gly\_Ser\_codes.pdf}\\
Text file containing the result of the computational analysis of the
network from S5.

\paragraph{Network - S7} \texttt{s7\_non-biological-networks.zip}\\
Network files of the analyzed combustion chemistries and the Martian
atmosphere chemistry.

\paragraph{Network - S8} \texttt{s8\_ntop.zip}\\
Network files of the artificial chemistry NTOP.

\paragraph{Text - S9} \texttt{s9\_ntop\_codes.pdf}\\
Text file containing the result of the algorithmic analysis of NTOP.

\section*{Acknowledgments}
We are grateful for many fruitful discussions with Stefan Artmann and
his valuable comments and suggestions. We thank
Marcel Hieckel for preparing the combustion chemistry data. 
We acknowledge funding by the DFG through the Jena School for Microbial Communication
(JSMC).

\bibliography{references}

\begin{landscape}

 \begin{table}[!p]
\scriptsize
 \caption{{\bf List of all analyzed systems stating their size, density,
   semantic capacity, the reference of the system, and the method used for
   analysis.}}
 \begin{tabular}{llllllllp{8cm}}
 Abbrev & \#species & \#reactions & \#closed sets & \#paths & $SC_{log}$ & Method  & Reference  & Description \\\hline

 FIG1A & 8  &  4   & 161 & 12  & 1 & c \& p   & this study & Network from Figure \ref{fig:examplenet}A.\\
 FIG1C & 6  &  4   & 41  & 16  & 1.58 & c \& p   & this study & Network from Figure \ref{fig:examplenet}C.\\
 GCMERGE  & 234 & 85 & n.a. & 170  & 4.09 & p & this study &
 Network reconstructed from the genetic codes reported at \cite{Elzanowski2010}\\[0.1cm]
 GCFULL   & 1364 & 1280  & n.a. & n.a  & 18.55 & t & this study &
 Theoretical estimate of $SC_{log}$ of a network, based on GCMERGE, generated by inserting all possible
 mappings between codons and amino acids\\[0.1cm]
 GCFULLSYNTSMALL & 16 & 8  & n.a.  & 200  & 2.32 & p & this study & Network with two codons, two amino acids, and synthetases.\\
 GCFULLSYNT   & 2,728 & 2,560 & n.a. & n.a.  & 20.55 & t & this study &
 Theoretical evaluation of a reaction network containing all possible
 mappings between the 64 codons and 20 amino acids with synthetases \\[0.1cm]
 MARS & 32 & 104 & 5,512 & $> 10^6$  & 0 & c & \cite{Nair1994}  &
				  Chemical processes occurring in the
				  Martian atmosphere during the daylight phase\\
 HYD  & 10 &  38 & 16    & $7.69 \cdot 10^4$   & 0 & c & \cite{Conaire2004}  & Combustion chemistry of hydrogen\\
 MET  & 37 & 340 & 4,136 & $> 10^6$  & 0 & c &  \cite{MethaneOxidation2001} & Combustion chemistry of methane\\
 ETH  & 57 & 752 & 5,136 & n.a.  & 0 & c &  \cite{Marinov1999} & Combustion chemistry of ethanol\\
 DME  & 79 & 708 & 8     & $> 10^6$  & 0 & c &\cite{Kaiser2000}  & Combustion chemistry of dimethyl ether\\
 NTOP & 16 & 207 &  244  & $474,218$  & 2.81 & c \& p &  \cite{ac:Ban93a} & Artificial chemistry based on
 binary strings operations \\[0.1cm]
 R\_NTOP   & 16 & 207 & 18.11 (s.e.=0.23)& n.a. &  0 (s.e.=0)& c & this study & Average
 on 1000 random networks of the same size and density as the NTOP network.\\
 RANDOM & varies &  varies   & varies & varies  & varies & c \& p
 & this study & Analysis of different random networks.\\\hline

 \end{tabular}
 \label{nbsystems}
 Abbrev.: c - closure based algorithm, p - pathway-based algorithm, t -
 theoretical analysis, $SC_{log} $ - logarithmized semantic capacity, n.a. -
 not available, s.e. - standard error of the mean\\
 $^*$: determined with $k=10000$.
 \end{table}
\end{landscape}
\end{document}